\documentclass[sigconf,10pt,a4]{acmart}
\vspace{1em}
\renewcommand\footnotetextcopyrightpermission[1]{} % removes footnote with conference information in first column
\pdfoutput=1

\usepackage{enumitem}
\usepackage{pgfplots,pgfplotstable}
\usetikzlibrary{patterns}
\usepackage{xurl}

\usepackage{balance}

\usepackage{hyperref}

\begin{document}

% ****************** TITLE ****************************************

\title{StreamChain: Rethinking Blockchain for Datacenters
	%Towards Sub-Millisecond Processing of Transactions in Blockchains
}

\author{Lucas Kuhring}
\affiliation{%
  \institution{IMDEA Software Institute}
}
\email{lucas.kuhring@imdea.org}

\author{Zsolt Istv\'{a}n}
\affiliation{%
  \institution{IMDEA Software Institute}
}
\email{zsolt.istvan@imdea.org}

\author{Alessandro Sorniotti}
\affiliation{%
	\institution{IBM Research - Zurich}
}
\email{aso@zurich.ibm.com}

\author{Marko Vukoli\'c}
\affiliation{%
  \institution{IBM Research - Zurich}
}
\email{mvu@zurich.ibm.com}

\begin{abstract}
Permissioned blockchains promise secure decentralized data management in business-to-business use-cases. In contrast to Bitcoin and similar public blockchains which rely on Proof-of-Work for consensus and are deployed on thousands of geo-distributed nodes, business-to-business use-cases (such as supply chain management and banking) require significantly fewer nodes, cheaper consensus, and are often deployed in datacenter-like environments with fast networking. However, permissioned blockchains often follow the architectural thinkining behind their WAN-oriented public relatives, which results in end-to-end latencies several orders of magnitude higher than necessary.

In this work, we propose \emph{StreamChain}, a permissioned blockchain design that eliminates blocks in favor of processing transactions in a \emph{streaming} fashion. This results in a drastically lower latency without reducing throughput or forfeiting reliability and security guarantees. To demonstrate the wide applicability of our design, we prototype StreamChain based on the Hyperledger Fabric, and show that it delivers latency two orders of magnitude lower than Fabric, while sustaining similar throughput. This performance makes StreamChain a potential alternative to traditional databases and, thanks to its streaming paradigm, enables further research around reducing latency through relying on modern hardware in datacenters.
\end{abstract}

\maketitle

\section{Introduction}

Blockchains (distributed ledgers) offer strong data integrity and reliability properties in untrusted environments that make them worth considering for distributed data management use cases beyond crypto-currencies. Their adoption, in particular that of \emph{permissioned} blockchains (in which membership is vetted, e.g., \cite{androulaki2018hyperledger}) is expected to bring benefits to enterprise consortia thanks to the logical centralization of datasets that previously resided at separate sites and were interconnected through various protocols. Prominent examples include financial use cases~\cite{Stellar}, supply-chain management~\cite{hackius2017blockchain} and provenance \cite{Ruan0DLOZ19}, to name but a few. 

As most enterprise software, permissioned blockchains are deployed predominantly in cloud environments, with all leading cloud providers such as  Amazon~\cite{AWSBlockchain}, IBM~\cite{IBP}, Microsoft~\cite{AzureBlockchain} and Oracle~\cite{OracleCloudBlockchain} offering hosted enterprise blockchain infrastructure and services. Even though this introduces the cloud provider as a centralization point, which is at odds with blockchain decentralization, in the long run, cloud providers are likely to offer ways of interconnecting blockchain nodes across multiple clouds while retaining high bandwidth and low latency communication. Such multi-cloud blockchain offerings, that span multiple cloud providers, are in fact already starting to appear~\cite{IBMBlockchainMulticloud}.  

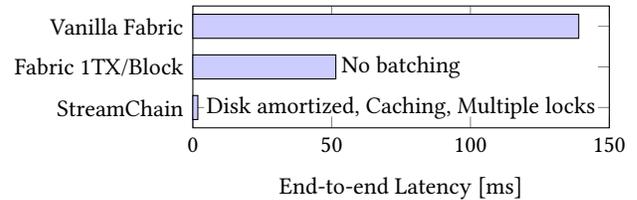
\begin{figure}[t]
	\begin{tikzpicture}
	\pgfplotstableread{ % Read the data into a table macro
		Label   Bandwidth   
		{StreamChain}      1.8
		{Fabric 1TX/Block}  51.5
		{Vanilla Fabric} 139
	}\datatable
	\begin{axis}[
	font=\small,
	width=0.4\textwidth,
	height=0.18\textwidth,
	xbar stacked,   % Stacked horizontal bars
	xmin=0,         % Start x axis at 0
	ytick=data,     % Use as many tick labels as y coordinates
	yticklabels from table={\datatable}{Label},  % Get the labels from the Label column of the \datatable
	xmin=0,
	xmax=150,
	ymin=-0.5,
	ymax=2.5,
	bar width=0.32cm,
	xlabel={End-to-end Latency [ms]},
	]
	\addplot [black, fill=blue!20] table [x=Bandwidth, y expr=\coordindex] {\datatable};    % Plot the "First" column against the data index
	\node[] at (axis cs: 75,1) {No batching};
	\node[] at (axis cs: 75,0) {Disk amortized, Caching, Multiple locks};
	\end{axis}
	\end{tikzpicture}
	\vspace{-1em}
	\caption{\label{fig:latency-comparison}The latency of Fabric is dominated by block-based batching and disk access. By removing batching and applying further optimizations, latency can be reduced by two orders of magnitude.}
\end{figure}

Even in cross-cloud enterprise blockchain deployments, geo-distribution is often neither required nor beneficial. For instance, in performance sensitive blockchain applications, such as stock exchanges~\cite{HongKong, Australia}, \emph{datacenter-like} deployments, characterized by  geographical proximity, are a better fit. Fault-diversity may be ensured by running nodes across different datacenters belonging to different providers, while geographical proximity allows network latencies to remain low and bandwidth high. 
%and whose characteristics are more akin single-datacenter deployments than WANs. 
%Therefore, ``datacenter-like'' deployments are not only the reality of today's permissioned blockchain, but also arguably their future. 

Unfortunately, when deployed in datacenter-like environments, all current permissioned distributed ledgers exhibit poor performance. For instance, they will take hundreds of milliseconds to commit transactions~\cite{androulaki2018hyperledger}, even if the networking latencies are in the order of microseconds (see Fig.~\ref{fig:latency-comparison}). This is because the design of most permissioned blockchains is based on that of public blockhcains and, as a result, they are often optimized for wide area networks. For instance, the batching of transactions into blocks is useful for amortizing the high computational cost of Proof-of-Work (PoW) consensus, as well as networking overheads; however, in permissioned blockchains, blocks are a source of added latency and should be removed.

In the light of emerging datacenter-like deployments, we introduce \emph{latency} as the first-class blockchain performance metric to complement throughput, which is generally perceived as the main metric of interest~\cite{Vukolic15}. We revisit the design decisions of permissioned blockchains from the perspective of reducing latency and, as a concrete instantiation of our design recommendations, we prototype \emph{StreamChain}, our \emph{streaming variant} of Hyperledger Fabric~\cite{androulaki2018hyperledger}. In a nutshell, StreamChain departs from block-based system design and chains transactions directly, in a streaming fashion, drastically reducing latency, but without jeopardizing security, reliability or throughput guarantees.

\smallskip

The contributions of this work are as follows:

\begin{itemize}

\item \emph{Inroducing a latency-efficient permissioned distributed ledger for datacenter-like environments:} By not batching transactions into blocks and by exploiting the parallelism of modern multicores, StreamChain can achieve end-to-end latencies close to 1.5\,ms without compromising throughput. As shown in Figure~\ref{fig:latency-comparison}, its latency is two orders of magnitude smaller than vanilla Fabric running on a local area network, and one  order of magnitude smaller than Fabric naively configured  to form single transaction blocks. Overall, StreamChain provides latencies that are comparable to traditional databases, thereby making it a more realistic alternative to consider.

\item \emph{Showcasing integration with modern datacenter accelerators:}  Even though consensus can be adapted for fast networks~\cite{istvan2016consensus,poke2015dare}, current permissioned ledgers do not benefit significantly from such optimizations due to the heavy use of batching. Conversely, latency improvements of consensus make a difference in StreamChain and, as proof-of-concept, we implemented an FPGA-based consensus service that makes it possible to reach commit latencies (i.e., latency excluding smart contract execution) below a millisecond.

\item \emph{Platform for future research:} StreamChain is an open-source\footnote{Source code available at: \url{https://gitlab.software.imdea.org/zistvan-public/streamchain} and \url{https://gitlab.software.imdea.org/zistvan-public/streamchain-benchmarks}} platform to explore throughput and latency improvements in permissioned ledgers. StreamChain builds on a long-term support (LTS) release of Fabric (v.1.4) which allows future  platform improvements to remain compatible with currently existing application ecosystems. 
\end{itemize}

This paper is organized as follows: we provide an overview of permissioned blockchains and Hyperledger Fabric in Section~\ref{sec:background}. The design and implementation of StreamChain is described in Section~\ref{sec:system}. We evaluate the system in Section~\ref{sec:evaluation} with a YCSB microbenchmark and a Supply Chain Management application. We talk about the next steps in Section~\ref{sec:future-work} and conclude in Section~\ref{sec:conclusion}.

\pagestyle{empty}
\section{Background and Related Work}
\label{sec:background}

\subsection{Permissioned Ledgers and Fabric}

Public ledgers do not authenticate nodes and instead they rely on PoW and similar consensus protocols to make Sybil attacks impractical. Permissioned (private) blockchains, on the other hand, authenticate all nodes and therefore allow the use of much cheaper consensus protocols. Nonetheless, permissioned ledgers inherit their design from public ones with many systems evolving from crypto-currency to more general use through ``smart contracts'' (e.g. Ethereum~\cite{wood2014ethereum}). Since the execution of the smart contracts has to be serial, many of these systems are severely limited in throughput when running complex contracts. As a result, Ethereum, for instance, adds a ``complexity'' fee to smart contracts in the form of ``gas'', thereby limiting  usability in enterprise scenarios.

\begin{figure*}[t]
\centering\includegraphics[width=0.75\linewidth]{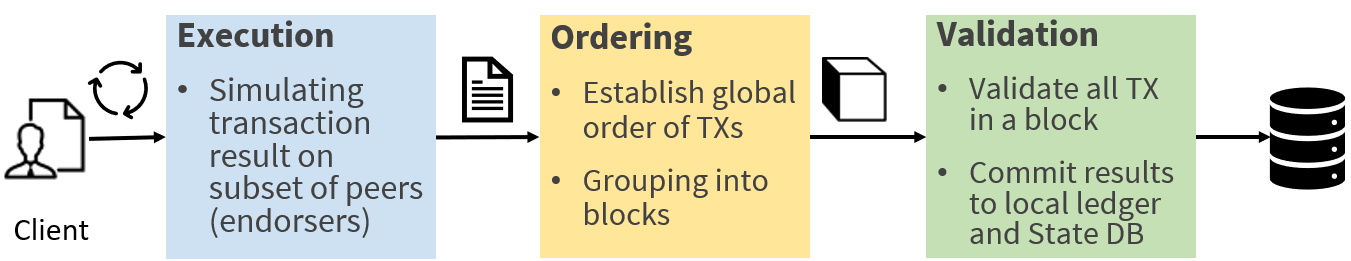}
\caption{\label{fig:fabric-processing}Fabric processes transactions in three steps, relying on two types of nodes: peers for the first and third and orderers for the second.}
\vspace{1em}
\end{figure*}

Hyperledger Fabric~\cite{androulaki2018hyperledger} is an open-source permissioned ledger that allows smart contracts (``chaincode'') to be developed in general purpose programming languages. It also removes the execution bottleneck by using an Execute-Order-Validate (EOV) approach (see Figure~\ref{fig:fabric-processing}). As opposed to the Order-Execute (OE) model used, for instance, in Ethereum, where smart contracts are ordered and then shipped to all nodes for execution, in EOV only a subset of nodes execute them. This is done in isolation on top of the ``materialized view'' of the ledger (Fabric's State DB) to determine their read/write sets and values (TX descriptors). The ordering nodes run a consensus protocol and establish the global order of these read/write sets which are then validated on all peers on their current State DB and committed to the ledger. The EOV model saves compute resources at peers and allows the execution of feature-rich smart contracts.

The drawback of the EOV model is that, since execution and committing happen concurrently on the nodes, it is possible that in the time it takes to order transactions, the underlying state of the variables in the read/write set have changed. This will result in a ``failed'' transaction whose results are not applied to the ledger state, even though its execution is recorded. This means that in case multiple clients execute chaincode that shares state, the successful transaction throughput (goodput) can be significantly lower than the raw throughput of the system.

%Most blockchain systems have the following in common: separation between role of consensus (agreeing on what goes into a block) and data management (persisting of ledger, running smart contracts, etc.).

%Systems such as Ethereum are limited in performance due to the sequential execution of contracts (OE model). 

%Hyperledger Fabric optimizes on this design by using EOV model, which reduces resource usage and allows for more parallelism. It, however, still largely inherits the same design.

%Highlight the failing-TX issue which comes from the fact that E sees stale data.

\subsection{Improvements and Alternative Designs}

Most related work sets out to increase the throughput of permissioned ledgers but end-to-end latencies of single transactions are typically not addressed. However, we believe that this is an important feature to improve, especially in use-cases where network latencies are less than in a wide area deployment. 

The authors of FastFabric~\cite{gorenflo2019fastfabric} propose various optimizations to Fabric which result in an unprecedented 20,000 ops/s throughput. They achieve this through a combination of multi-threaded batch processing, changes to the ordering service and caching deserialized data structures in the validation phase. We have adopted the latter technique in StreamChain and reduced, as a result, the cost of the validation step by 10\% (see details in Section~\ref{sec:batchandcache}).

Sharma et al.~\cite{sharma2019blurring} focus on increasing goodput in the EOV model and demonstrate that, by relying on database techniques for  concurrency control, it is possible to reorder transactions within a batch inside the ordering service in a way that minimizes the number of failing transactions. This idea is applicable to our case as well but it changes an assumption that Fabric makes about the ordering service, namely, that the contents of the transactions are opaque. Depending on the trust model, this raises concerns related to malicious reordering.

CAPER~\cite{amiri2019caper} demonstrates how multiple applications that share the same distributed ledger can leverage the fact that they operate on disjoint parts of the dataset in order to increase the overall throughput. This is achieved by separating the ordering of operations inside an application from that of ordering across applications. The authors propose a supply chain management workload for benchmarking, similar to the one we use for evaluation.

There is an emerging class of distributed ledgers that do not incorporate the notion of blocks at all and operate on a per-transaction basis. Two well known examples are Corda~\cite{corda} and IOTA~\cite{popovtangle}, which were designed with a very specific financial services use-case in mind. Instead of storing data in a single chain, a directed acyclic graph (DAG) was used. This allows executing transactions between various subsets of the peers without the burden of global ordering, but also results in systems with different properties than traditional permissioned ledgers.

\subsection{Preliminary Version of This Work}

In a preliminary version of this work~\cite{istvan2018streamchain} we demonstrated that the main idea behind StreamChain is viable by using a mock implementation that consisted of running Fabric v1.0 configured with 1TX per block and to run in main-memory (i.e., without persistence). In this work we provide a full implementation and further optimizations by:
\begin{itemize}
	\item re-implementing StreamChain based on Hyperledger Fabric~1.4 LTS in a way that produces a transaction log that remains compatible with vanilla Fabric; 
	\item introducing additional pipelining in the validation stage to increase throughput; 
	\item providing a tunable disk write batching mechanism that ensures persistence while hiding I/O cost; and
	\item showing that the streaming execution model allows us to further optimize the ordering service, for instance, by including specialized hardware based on FPGAs. 
\end{itemize}

Even though the throughput of StreamChain could be further increased, it is already a feasible alternative to traditional databases in emerging use-cases, even for contention-heavy workloads. Furthermore, StreamChain opens up novel opportunities for incorporating specialized hardware, such as FPGAs or SmartNICs, at various levels since the time spent within each processing step is directly exposed.

\section{Our solution: StreamChain}
\label{sec:system}

\subsection{Which Scenarios is It For?}

StreamChain targets datacenter-like deployments that are closer to a single datacenter  than to geo-distributed environements and offer high bandwidth and low latency networking across nodes. Given that, at the moment, virtually all hosted permissioned ledgers are single-cloud, in the current prototype  StreamChain tolerates Byzantine peers but uses crash fault tolerant (CFT) ordering, just like Fabric v1.4 LTS. This means that the participants have to trust ordering nodes not to be malicious/Byzantine. In the long run, with the emergence of multi-cloud deployments, it will be necessary to integrate a Byzantine fault tolerant (BFT) ordering service (e.g., \cite{SousaBV18}) with our streaming optimizations. We discuss the path to achieving this in Section~\ref{sec:future-work}.

\subsection{Insights and System Design}

StreamChain builds on the following insight: in permissioned ledgers that have fast consensus, blocks have a negative effect on response times in low latency networks. Removing blocks, however, does not change the total amount of computation in the system. It still requires careful parallelization and pipelining of cryptographic operations and disk accesses in order to achieve reasonable throughput. An important side-effect of reducing latency in EOV systems is that it results in higher goodput: lower commit latencies mean less stale data for endorsement, which in turn yields less failed transactions in validation.

\subsubsection{Batching and Caching}
\label{sec:batchandcache}

In Fabric, both the ordering and the validation (commit) stages are writing to disk for each block that is processed in the system, and all transactions, even those that only read from the state, have to be persisted in the ledger. If we would naively set the size of blocks to a single transaction then latencies would be dominated by the disk flushing cost (see Figure~\ref{fig:latency-comparison}) and throughput would be reduced to the IOPS of the underlying device. 

In StreamChain, we replace the code-path that flushes the ledger to disk with a local batcher that waits until enough data has accumulated or a time-out is reached (e.g., 100ms) before flushing to disk in an asynchronous manner. To avoid inconsistencies, all buffers are flushed immediately if the code on top performs a read of the underlying data. The same approach applies to all data structures in the peers that are persisted to disk (ledger state, index structures, etc.), with the exception of the materialized view of the ledger that is stored in the State DB. 

In Fabric, LevelDB is used by default for the State DB and it persists its own data. The main reason for writing to disk in LevelDB is to speed up recovery in case a node fails and restarts, but we argue that it would be enough to checkpoint the key-value store off the critical path. Hence, we configure LevelDB to run in main-memory. Since the ledger is still persisted to disk, durability is not compromised.

Fabric uses gRPC for the communication between participants of the network and it uses Protocol Buffers to marshal and unmarshal messages. This can be an expensive operation with the block data structure that the orderers send to the peers. Unmarshalling happens selectively and at different levels of granularity; since parts of the block are used several times during the validation phase this leads to inefficiencies. The data structure we use to wrap transactions in StreamChain are very similar to blocks with less overall metadata, and therefore this inefficiency applies to StreamChain as well. As a solution, we implement an idea from FastFabric~\cite{gorenflo2019fastfabric}, namely, a cache that stores unmarshalled block contents to avoid repeated work. Since transactions are received on a single thread and never modified inside the peer, the cache doesn't need to be protected by locks against concurrency.

Beyond the above optimizations, we also modified the way in which the peer reads in its configuration parameters, caching them for subsequent accesses. Furthermore, we disabled an auxiliary data structure, called the history DB that exists to support data provenance queries. Nonetheless, the same batching techniques apply to this data structure as to the main ledger structures.

\subsubsection{Using Available Parallelism}

The peers receive transactions (or blocks in the case of Fabric) from the ordering nodes over the network. These transactions are passed to the Validation logic that performs signature and read/write set checks before it records them in the ledger. In StreamChain we divided the Validation logic into 3 pipeline stages to take advantage of multi-core machines. This layout extends Fabric's two-stage pipeline by one stage and, whereas in Fabric parallel signature checks are only carried out within a block, in StreamChain they work on incoming transactions directly. The pipeline is organized as follows:

\begin{enumerate}[leftmargin=*]
\vspace{-0.25em}
  \item \textbf{Message authentication and endorsement signature check} -- this step is the most expensive one because it requires checking various signatures to decide whether each transaction has been properly endorsed and ordered. Since there are no data dependencies across transactions, we rely on several cores to parallelize this check and collect the results in a FIFO order. In principle all idle cores of the CPU could be used for this purpose, but we found that in our experimental setup 6-8 were enough to result in pipeline stages with almost equal processing times.
%  \vspace{-0.35em}

  \item \textbf{Read/write set check and ledger commit} -- this step iteratively checks the transactions' read/write sets against the current ledger state (materialized in the State DB) and if no conflicts are found, it commits them to the ledger and updates the State DB. This operation needs to be performed sequentially for transactions so it runs in a single thread. Fabric uses a single reader/writer lock around the State DB and limits parallelism in our pipelined version between endorsement and validation. Therefore, we replace the single lock with a series of locks that allow more fine-grained access. 
%\vspace{-0.35em}

  \item \textbf{Additional housekeeping} -- after updating the ledger state, various additional operations are performed on auxiliary data structures used, for instance, in the gossip protocol. Even though we disable gossip in our deployment  (as it's not needed in deployments with plenty of networking bandwidth between ordering nodes and peers), it is important that StreamChain doesn't remove Fabric features. Therefore, these data structure updates are still performed but in their own pipeline stage, which moves them off the critical path.
%  \vspace{-0.25em}
\end{enumerate}

\section{Exploring New Opportunities}

Since StreamChain exposes the cost of each execution step, it acts as a platform for future exploration and lets us evaluate the usefulness of various emerging hardware features in the datacenter. In this work we focus on the ordering step and show that by incorporating Field Programmable Gate Arrays (FPGAs), which are increasingly available in datacenters, into the ordering service we are able to reduce its latency further and allow it to scale to higher throughput in the future. We target a scenario where the permissioned ledger is hosted by a service provider that has existing infrastructure (e.g. Microsoft Catapult) that can be used to improve the overall performance of the blockchain.

\subsection{FPGAs in the Datacenter}

Field programmable gate arrays (FPGAs) are hardware chips that are composed of a collection of small look-up tables (LUTs), on-chip memory (BRAM) and specialized digital signal processing units (DSPs), which can be configured and interconnected to implement any hardware circuit. In comparison to traditional processors, FPGAs allow for fine-grained dataflow parallelism due to the fact that all ``code'' is executing in parallel in the device. The energy footprint of FPGAs is an order of magnitude lower than that of server-grade CPUs (and even though it is higher than that of ASICs, FPGAs can be reprogrammed freely, whereas ASICs have fixed functionality). For a more detailed description of FPGA internals and a summary of their strengths in data processing scenarios, we direct the reader to the book by Teubner and Woods~\cite{teubner2013book}.

FPGAs are being explored in cloud context both for compute intensive tasks related to machine learning~\cite{chung2018serving} and as low latency key-value store solutions~\cite{xu2016bluecache,fukuda2014caching,istvan2017caribou} mainly due to their ability to provide predictable, line-rate behavior even when performing near-data processing tasks. 

Importantly, they are already being used in production for infrastructure acceleration, for instance in Project Catapult~\cite{firestone2018azure}, to offload networking tasks in Azure virtual machines. With the general availability of FPGAs in the cloud (the number Catapult-enabled Azure servers in 2018 was reported to be more than a million~\cite{firestone2018azure}), they could be used to provide functionality packaged as a service with a lower energy footprint, hence lower cost, and more predictable performance than software-based solutions.

\begin{figure}[t]
\includegraphics[width=\linewidth]{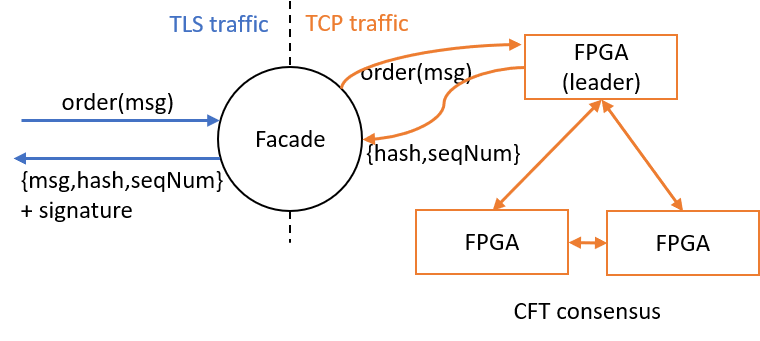}
\caption{\label{fig:fpga-ordering}We integrate a crash fault tolerant ordering service built with FPGAs into StreamChain by using a software node to act as facade between the service and the blockchain peers.}
\vspace{-0.5em}
\end{figure}

\subsection{FPGA-based Ordering}

Fabric is designed such that the ordering service can easily be replaced with custom implementations. In version 1.4, it has a service built with Raft~\cite{ongaro2014search} as its main way of ordering blocks/transactions. As an alternative, an Apache Kafka based one is available but this introduces higher latencies and more complexity to the system. Fabric also offers a ``solo'' orderer comprising of a single node (hence no fault tolerance) that is useful for development and testing purposes. 

For the FPGA-based ordering service, we build upon our earlier work on CFT consensus on stand-alone FPGAs~\cite{istvan2016consensus}. That work implements Zookeeper's Atomic Broadcast protocol to replicate write operations in a key-value store running on the same nodes. One of the main benefits of using FPGAs in this context is that they can achieve 10Gbps network-bound performance similar to RDMA-based systems without specializing the network protocol, relying instead on commodity networks and TCP sockets for communication across the consensus nodes and the clients. 

For this work, we modified the interface of the nodes to expose the following three operations: 1)~\emph{order}, that takes a transaction (a BLOB), replicates it across nodes and returns the sequence number assigned to the transaction, 2)~\emph{get}, to retrieve the transaction ordered with a specific sequence number and 3)~\emph{getLast}, to retrieve the latest ordered transaction and its sequence number. To achieve this functionality, we modified the key-value store implementation on the FPGA to expose the sequence numbers of the consensus algorithm that were hidden before. The design does not rely on blocks and computes the SHA256 hash of each individual transaction inside the FPGA. This hash value is appended to the transactions before they are stored and is returned with \emph{get}s, thereby forming  the ``links'' of the chain. 

Our prototyping boards do not have persistent storage, therefore we explored the feasibility of using NVDIMMs for providing durability in the future. We added a module in front of the memory controller on the FPGA to simulate the timings and bandwidth of Intel's Optane NVDIMMs~\cite{izraelevitz2019basic}. Given the throughput levels of StreamChain, using lower bandwidth NVMe Flash could also be an option.

For integration with the rest of the nodes in StreamChain we rely on the code from the ``solo'' orderer to act as a facade between peers and FPGAs (Figure~\ref{fig:fpga-ordering}). It deals with aspects such as TLS flow termination, management messages, etc., and unwraps the transactions and submits them to the FPGA nodes for ordering. For our prototype we used a single facade, but since it is stateless, it could be deployed on multiple nodes to help with fast failover.

\section{Evaluation}
\label{sec:evaluation}

\subsection{Benchmarking Applications}

In the Evaluation of StreamChain, we focus on answering the following questions:
\begin{itemize}

  \item Can low latency be achieved simply by using blocks of a single transaction?
  \item What is the relative cost of the different steps and pipeline stages?
  \item What is the throughput of StreamChain and Fabric with more complex chaincodes?
  \item Do failing transactions reduce the useful throughput (goodput) of the system significantly?
  \item Could StreamChain compete with a database?

\end{itemize}

\noindent
To answer these questions, we designed two different benchmarks (YCSB and SCM), as described in the following.

\subsubsection{YCSB-Like Microbenchmark}

As a micro-benchmark we relied on the YCSB suite to generate 1KB key-value operations and encoded these as invocations of chaincodes that insert/update/read a value belonging to a key. Experiments were driven by a peer executable that connects to endorsers and invokes the chaincode directly, avoiding an additional RPC from the original Java-based YCSB client. We noticed that, since in the original Fabric version most time is spent on disk writes to update the ledger state, the actual type of operation (set or get) didn't play an important role for performance. We used two workloads: 90\% inserts/10\% reads (YCSB-90) and equal read/update ratios (YCSB-50).

\begin{figure}[t]
\includegraphics[width=\linewidth]{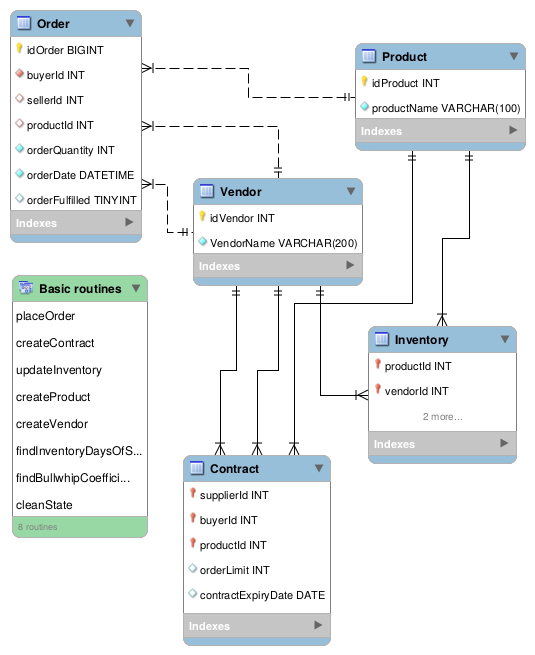}
\caption{\label{fig:mysql-schema}The SCM example implemented in MySQL consists of tables describing relationships between vendors and keeps track of their orders and current inventories. Both transactional and analytical queries are implemented as stored procedures.}
\vspace{-0.5em}
\end{figure}

\subsubsection{Supply Chain Management Scenario}

Since supply chain management (SCM) is an important use-case for distributed ledgers, we designed a benchmark targeting this scenario: vendors order products from each other and update their inventories, provided that they have put a contract in place (n.b. this is not a smart contract, but a condition that allows ordering a specific type of product). The corresponding schema is shown in Figure~\ref{fig:mysql-schema}. In addition to transactional queries, such as \emph{create contract}, \emph{place order}, \emph{update order}, etc., two analytical queries are provided: one to compute the \emph{days of supply (local analytics)} and one to compute the \emph{bullwhip coefficient (global analytics)}. The former determines the number of days for which a  vendor can supply a product based on current demand and inventory; the latter calculates a number that represents the overall variation between demand and supply across the supply-chain~\cite{fransoo2000measuring}.

We implemented all operations as stored procedures in MySQL 8.0 and ran the benchmark from a multi-threaded Java application issuing requests in a closed loop over JDBC. For the distributed ledger version we implemented the queries as chaincode written in Go and organized the key-space by the indexes and primary keys defined in the SQL schema. Data is encoded as JSON within the ledger's key-value pairs. For running the benchmark on Fabric and StreamChain, we used the Java application to export chaincode invocations and we used the same peer executable as for the microbenchmark.

In the supply chain management scenario we focused on the case where contention happens with high probability and, as a result, failing transactions could reduce goodput. To this end, we ran the benchmark with 50 vendors selling 500 products with 6000 contracts between them. Each order issued as part of the benchmark was based on one of these contracts. We varied the portion of transactional queries between 95\% (SCM-95) and 99\% percent (SCM-99).

\subsection{Deployment}

We ran the experiments on a local 10Gbps cluster of eight servers with Intel Xeon E-2186G CPUs (6x3.80GHz), 32GB of RAM and regular HDDs. The blockchain network was composed of 5 peers and 3 ordering service nodes for Raft, respectively 1 facade node for the FPGA ordering version with three Xilinx VCU1525 FPGA boards connected to the same switch. We ran MySQL Version 8.0.15 on one of the servers. For benchmarking we used a single client machine with the two workloads described in the previous section and issued 10000 operations not counting data loading, warm-up and cool-down phases (first and last 10\% of operations). For benchmarking MySQL, the client was situated on the same machine. Unless otherwise stated we used Fabric with the default block batching of 10 transactions as a baseline.

\subsection{Latency and Throughput}

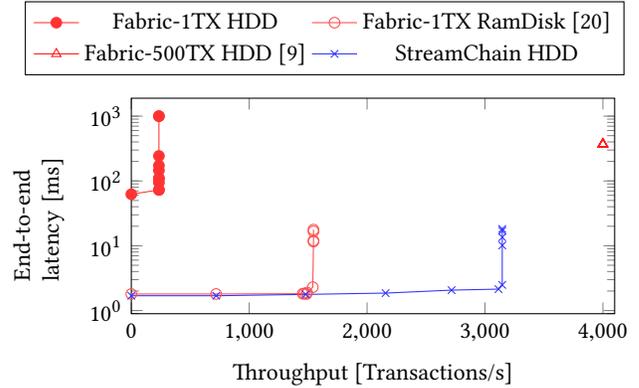
\begin{figure}[t]
  \begin{tikzpicture}  
  \pgfplotstableread{ % Read the data into a table macro
tputSC  latSC  tputMem  latMem  tputVan latVan tputES latES
0 1.691982  0 1.800272  0 62.398235 4000  371
720.464035  1.691982  719.127489  1.800272  234.29982 72.729274 4000  371
1479.62642  1.774207  1453.817138 1.819943  234.29982 72.729274 4000  371
2157.084334 1.859535  1467.27778  1.819951  234.29982 94.227078 4000  371
2716.630412 2.064765  1486.551065 1.800427  234.29982 107.313553 4000  371
3114.709001 2.140784  1490.908369 1.893789  234.29982 111.126706 4000  371
3146.177234 2.488911  1539.357454 2.293441  234.29982 145.197615 4000  371
3146.646727 10.122387 1545.248918 11.602079 234.29982 172.772384 4000  371
3146.646727 13.517588 1545.248918 12.013037 234.29982 243.166697 4000  371
3146.646727 17.023114 1545.248918 16.664926 234.29982 999 4000  371
3146.646727 18.173327 1545.248918 17.744974 234.29982 999 4000  371
  }\datatable
  \begin{axis}[
   legend style={at={(0.4,1.45)},
    anchor=north},
  legend columns=2,
  font=\small,
  width=0.45\textwidth,
  height=0.25\textwidth,
  ylabel={End-to-end latency [ms]},
  ylabel style={text width=2cm},
  xlabel={Throughput [Transactions/s]},
  ymode=log,
  xmin=0,
  xmax=4100,
  ]
  \addplot [red!80, mark=*] table [y=latVan, x=tputVan] {\datatable};  
  \addplot [red!70, mark=o] table [y=latMem, x=tputMem] {\datatable};    
  \addplot [red!90, mark=triangle] table [y=latES, x=tputES] {\datatable};  
  \addplot [blue!80, mark=x] table [y=latSC, x=tputSC] {\datatable};  

  \legend{Fabric-1TX HDD, Fabric-1TX RamDisk~\cite{istvan2018streamchain}, Fabric-500TX HDD~\cite{androulaki2018hyperledger}, StreamChain HDD}

  \end{axis}
  \end{tikzpicture}
\caption{\label{fig:latencytput-comparison} Fabric with 1TX/Block lowers response times at the cost of throughput. StreamChain offers as low a response time as Fabric run on top of a RamDisk, but 2x better throughput.}
\end{figure}

\textbf{Can low latency be achieved simply by using blocks of a single transaction?}
To demonstrate that naively setting the block size of Fabric to one does not result in great improvements, and hence the optimizations in StreamChain are needed, we show latency as a function of throughput in Figure~\ref{fig:latencytput-comparison} when using the YCSB-90 workload (we also compare to Fabric in~\cite{androulaki2018hyperledger}, batching 500TXs on average, that uses a comparable micro benchmark). If we set the block size to one (1TX), disk access will dominate both latency and throughput. If, in addition, we ``remove'' the disk overhead by running on a RamDisk (as we did in our preliminary work~\cite{istvan2018streamchain}), latency can be lowered significantly but the system becomes non-persistent. In StreamChain we can maintain the low latency behavior and keep persistence, while achieving higher throughput. 

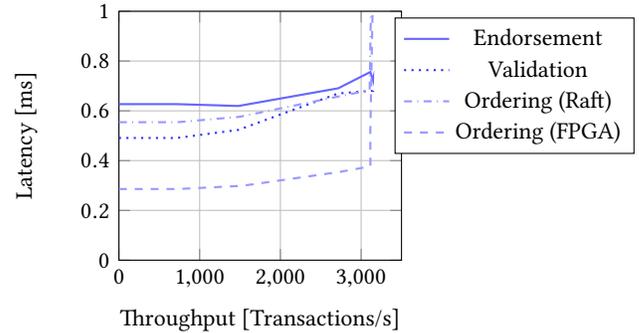
\begin{figure}[t]
  \begin{tikzpicture}  
  \pgfplotstableread{ % Read the data into a table macro
tput  end valid rafttput  raftlat fpgatput  fpgalat
0 0.626896  0.490991  0 0.554353  0 0.285909
720.464035  0.626896  0.490991  727.747835  0.554353  735.60972 0.285909
1479.62642  0.619457  0.523429  1494.829731 0.576035  1525.524639 0.298717
2716.630412 0.690574  0.669816  2713.596988 0.657231  2737.085947 0.354629
3114.709001 0.756105  0.680135  3105.218516 0.683009  3114.954583 0.376964
3146.177234 0.723325  0.680135  3122.605898 1.032682  3113.148865 0.50425
3146.646727 0.728017  0.680135  3146.646727 8.887312  3146.646727 1.090025
3140.007982 0.719912  0.695386  3140.007982 11.755566 3140.007982 1.559494
  }\datatable
  \begin{axis}[
  legend style={at={(1.4,0.975)},
    anchor=north,},
   legend columns=1,
  font=\small,
  width=0.3\textwidth,
  height=0.275\textwidth,
  ylabel={Latency [ms]},
  xlabel={Throughput [Transactions/s]},
  grid=both,
  %ymode=log,
  ymax=1,
  ymin=0,
  xmin=0,
  xmax=3500,
  ]
  \addplot [blue!60, style={line width=0.8pt}] table [y=end, x=tput] {\datatable};  
  \addplot [blue!80, dotted, style={line width=0.8pt}] table [y=valid, x=tput] {\datatable};    
  \addplot [blue!40, dashdotted, style={line width=0.8pt}] table [y=raftlat, x=rafttput] {\datatable};  
  \addplot [blue!40, dashed, style={line width=0.8pt}] table [y=fpgalat, x=fpgatput] {\datatable};    

  \legend{Endorsement, Validation, Ordering (Raft), Ordering (FPGA)}

  \end{axis}
  \end{tikzpicture}
\caption{\label{fig:latency-breakdown}The main components of the StreamChain pipeline have all predictable latency behavior. The FPGA-based ordering service cuts latency in half when compared to the Raft-based one.}
\end{figure}

\textbf{What is the relative cost of the different steps and pipeline stages?}
The latencies inside StreamChain are reduced and balanced across the three stages of the EOV model. Figure~\ref{fig:latency-breakdown} shows the breakdown of latency as well as its evolution with increasing load. When using FPGAs to run the ordering service instead of Raft, half of its latency (0.3\,ms) can be saved. A further reduction of latency would be possible by replacing the facade node with in-FPGA TLS termination. Overall, commit latencies (O+V) are below 1\,ms and end-to-end latencies (E+O+V) under 1.5\,ms up to 2000TX/s.

\subsection{SCM Benchmark}

\textbf{What is the throughput with more complex chaincodes?}
The SCM workload allows us to measure the latency of the systems with more complex operations. \mbox{Table~\ref{fig:latency-scm} shows} the end-to-end latencies of a single client. In the case of Fabric and StreamChain, the latency decreases when the fraction of transactional queries is increased from 95\% to 99\%. This is expected because analytical queries take longer than transactional ones and often keep a lock on the entire key-space (e.g., in the case of the Bullwhip coefficient) while being endorsed, thereby slowing all validations down. The database exhibits the opposite behavior, whereby each transactional query incurs several write disk accesses, but analytics can be performed on the buffered pages. While there certainly are databases that offer better performance than MySQL, overall, StreamChain delivers competitive latencies.

\begin{figure}[t]
  \begin{tikzpicture}  
  \pgfplotstableread{ % Read the data into a table macro
f10rel  f10 f100rel f100  sc10rel sc10  sc100rel  sc100
0 0 0 0 0 0 0 0
0.431772836 0.04  0.4335955 0 0.220187631 0.04  0.219871708 0
0.869222095 0.07  0.875730602 0 0.45732161  0.06  0.451916108 0.01
0.945580506 30.59 0.963332579 5.54  0.826988761 0.35  0.825099738 0.02
0.95593233  33.39 0.973739404 6.68  0.978379687 1.13  0.980615012 0.13
0.995064774 37.39 1.000677496 8.95  0.98890714  29  0.988111705 5.55
1 37.26 1 8.97  1 32.26 1 6.34
  }\datatable
  \begin{axis}[
  legend style={at={(0.325,1.1)},
    anchor=north,},
  font=\small,
  width=0.45\textwidth,
  height=0.275\textwidth,
  ylabel style={text width=2cm},
  ylabel={Percentage of failing TX [\%]},
  xlabel={Fraction of maximum throughput},
  grid=both,
  %ymode=log,
  ymax=40,
  xmin=0.7,
  xmax=1,
  ]
  \addplot [red!80, mark=o, style={line width=0.5pt}] table [y=f10, x=f10rel] {\datatable};    
  \addplot [blue!80, mark=triangle, style={line width=0.5pt}] table [y=sc10, x=sc10rel] {\datatable};  
  \addplot [red!80, mark=*, style={line width=0.5pt}] table [y=f100, x=f100rel] {\datatable};    
  \addplot [blue!80, mark=triangle*, style={line width=0.5pt}] table [y=sc100, x=sc100rel] {\datatable};    

  \legend{Fabric 1k keys, StreamChain 1k keys, Fabric 10k keys,  StreamChain 10k keys}

  \end{axis}
  \end{tikzpicture}
\caption{\label{fig:goodput-ycsb}When running the YCSB-50 microbenchmark, transactions start failing sooner in Fabric than in StreamChain, which retains almost 100\% goodput close to maximum throughput (1700TX/s, resp. 3200 TX/s). The difference in behavior is even more visible with a 1000 key working set.}
\end{figure}
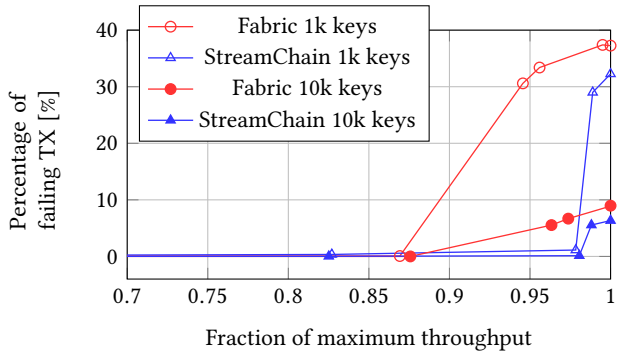

\textbf{How much do failing transactions reduce the useful throughput (goodput)?}
We explore the effect of small working set sizes (more read-after-write contention) on the goodput of StreamChain using the YCSB-50 workload with equal updates and reads and pick the working set as 100\%, respectively 10\%, of the 10k key space. As shown in Figure~\ref{fig:goodput-ycsb}, Fabric has a negligible number of failing transactions until it reaches 87\% of its maximum throughput, where the percentage of failing transactions quickly rises to 40\%. Failing transactions are not reissued in this benchmark, but if they were, depending on the backoff policy it could happen that the system is not able to make meaningful progress at this point. Thanks to the reduced data staleness, StreamChain shows a more graceful degradation that happens much later (beyond 97\% of its maximum throughput). Even though it cannot eliminate the problem entirely, StreamChain can operate close to saturation with virtually no failing transactions as soon as workloads don't have very small ``hotspots''.

\begin{table}[t]
\caption{\label{fig:latency-scm}The SCM benchmark provides a real-world estimate of end-to-end latencies since each query/chaincode performs several reads and writes.}
%\small
\begin{center}
\begin{tabular}{ |c|c|c| } 
 \hline
System  & SCM-95 & SCM-99 \\ 
\hline
Fabric & 296.3 ms & 169.6 ms  \\
StreamChain & 14.8 ms   & 4.4 ms  \\ 
MySQL & 36.3 ms   & 37.5 ms  \\ 
 \hline
\end{tabular}
\end{center}
\vspace{-0.5em}
\end{table}

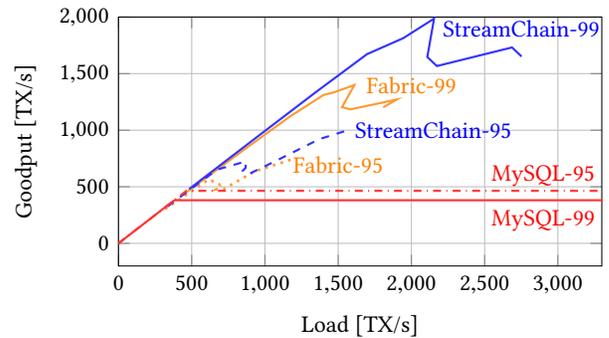
\begin{figure}[t]
  \begin{tikzpicture}  
  \pgfplotstableread{ % Read the data into a table macro
sc99tput  sc99gp  sc95tput  sc95gp  fab99tput fab99gp fab95tput fab95gp db99tput  db95tput  ram99tput ram95tput
482.292831  479.736679  333.120513  324.4260676 442.280423  430.9138161 318.252445  302.3398228 381 464 6989  2943
946.400709  940.3437445 671.501117  651.2217833 865.488972  835.2834069 616.643774  564.7223682 381 464 6989  2943
1344.476874 1333.452164 868.800093  723.7104775 1175.32387  1125.137541 685.631114  524.5078022 381 464 6989  2943
1697.033889 1672.087491 868.16234 667.0091258 1400.091801 1311.325981 692.293519  508.4895897 381 464 6989  2943
1944.575431 1813.511047 848.987531  642.7684597 1466.384508 1332.796879 662.892022  475.8901826 381 464 6989  2943
2154.706997 1986.20891  875.11979 623.6103624 1610.658114 1398.695506 700.853343  483.3084653 381 464 6989  2943
2108.580492 1646.801364 886.652777  610.460437  1525.731548 1203.344472 714.776942  477.8998634 381 464 6989  2943
2173.811423 1566.883274 1067.386009 723.6877141 1582.888153 1184.158627 931.513406  636.8757157 381 464 6989  2943
2686.656422 1731.281398 1370.286068 920.4211519 1898.885588 1270.544347 1084.391173 700.6251369 381 464 6989  2943
2752.359917 1651.41595  1562.864538 998.6704398 1807.793644 1202.001994 1187.520599 743.506647  381 464 6989  2943
  }\datatable
  \begin{axis}[
  legend style={at={(1.45,0.95)},
    anchor=north},
  font=\small,
  width=0.45\textwidth,
  height=0.275\textwidth,
  ylabel={Goodput [TX/s]},
  xlabel={Load [TX/s]},
  grid=both,
  %ymode=log,
  xmin=0,
  xmax=3300,
  ymax=2000
  ]
  \addplot [orange!80, style={line width=0.8pt}] table [y=fab99gp, x=fab99tput] {\datatable};  
  \addplot [orange!80, dotted, style={line width=1pt}] table [y=fab95gp, x=fab95tput] {\datatable};    
  \addplot [blue!80, style={line width=0.8pt}] table [y=sc99gp, x=sc99tput] {\datatable};  
  \addplot [blue!80, dashed, style={line width=0.8pt}] table [y=sc95gp, x=sc95tput] {\datatable};    

  \addplot[red!80, style={line width=0.8pt}] coordinates {(0,0) (381,381) (3600,381)};
  \addplot[red!80, dashdotted, style={line width=0.8pt}] coordinates {(0,0) (464, 464) (3600,464)};

  \node[red] at (axis cs: 2900,600) {MySQL-95};
  \node[red] at (axis cs: 2900,200) {MySQL-99};

  \node[blue] at (axis cs: 2750,1900) {StreamChain-99};  
  \node[blue] at (axis cs: 2150,1000) {StreamChain-95};

  \node[orange] at (axis cs: 2000,1400) {Fabric-99};
  \node[orange] at (axis cs: 1500,700) {Fabric-95};

  %\legend{Fabric-99, Fabric-95, StreamChain-99, StreamChain-95, MySQL-99, MySQL-95}

  \end{axis}
  \end{tikzpicture}
  \vspace{-1em}
\caption{\label{fig:goodput-scm}StreamChain delivers higher goodput than Fabric for the SCM use-case, even though failed transactions become more common with increased load and a higher percentage of analytical queries.}
\end{figure}

We also measure goodput with the SCM benchmark. The results, in Figure~\ref{fig:goodput-scm}, show two trends: First, by increasing the percentage of transactional queries (from 95\% to 99\%), the blockchain solutions become faster. Second, both Fabric and StreamChain have a drop in goodput sooner than with the YCSB benchmark. Both trends are to be expected. The endorsment of the analytical operations takes longer and requires locking a large portion of the key-space, thereby slowing down validation. Furthermore, concurrent updates to the inventory of vendors result in failing validations, thereby lowering goodput. StreamChain, however, consistently outperforms Fabric. 

\textbf{Could StreamChain compete with a database?}
when executing the SCM benchmark, MySQL delivers somewhat lower throughput than StreamChain, but of course no failing transactions (in Figure~\ref{fig:goodput-scm} we artificially extend the lines to the right once MySQL reaches maximum throughput). The database is limited by high lock contention and disk flushes for the transactional queries. If we run MySQL on a RamDisk, removing the disk overhead, its throughput increases significantly, to 3000TX/s for SCM-95 and 7000TX/s for SCM-99. These numbers can be regarded as an upper bound for traditional database throughput. Overall, StreamChain can deliver throughput within the range defined by these lower and upper-bounds on a single machine even though it executes in a distributed manner.

%\begin{figure}[t]
%\includegraphics[width=\linewidth]{figs/scm-latency.png}
%\caption{\label{fig:latency-scm}The SCM benchmark provides a real-world estimate of end-to-end latencies since each transaction performs several reads and writes. Overall, however, the latency savings remain and are visible when compared to Fabric.}
%\end{figure}

%MySQL   tx_po   a_doi   a_bwc   tput  a_bwc_Range
%1       37.9    0.49    5.27    29    2..8
%2       70.83   0.48    5.78    51
%4       80.56   0.78    8.57    99
%8       83.19   2.12    15.8    152

\section{Future Work}
\label{sec:future-work}

\subsection{BFT Ordering}
In order to reduce the necessary trust in a service provider and to ensure that permissioned distributed ledgers can be run in multi-cloud deployments, it is important to have a Byzantine fault tolerant (BFT) ordering service implementation. Using a BFT consensus protocol would also make the system more resilient to arbitrary failures. 

Even though there is already work providing BFT ordering in Fabric~\cite{SousaBV18}, integration with StreamChain will require modifications to the BFT implementation. This is because batching, used as a default in such protocols~\cite{bessani2014state,SousaBV18,yin2019hotstuff}, has to be eliminated without reducing throughput to impractical levels. An additional challenge is that, since each peer has to be connected to a majority of the BFT ordering nodes to accept transactions as ordered, the jitter across ordering nodes has to be minimized without reducing their ability to sustain high network bandwidth. There are no BFT implementations that fulfill all of the above requirements at the same time because traditionally they have been optimized for geo-distributed operation. Nonetheless, even though challenging, we believe that it is possible to build such an ordering service if we rely on low latency networks and using hardware accelerators both for cryptographic operations (signing, verifying, etc.) and data movement operations (marshaling and unmarshaling) withing the protocols.  

\subsection{Concurrent State Access}

Currently, endorsement and validation have to concurrently access the State DB inside endorsing peers. This limits throughput and can increase latencies when locking large portions of the key-space. For this reason, concurrency-enabling optimizations of the State DB will yield the biggest immediate benefits in StreamChain. Of course, fundamentally, the problem of failed transactions cannot be eliminated by changing the State DB. Instead, research is required into a hybrid between the OE and EOV execution models that allows more flexibility in post-order validation.

\section{Conclusions}
\label{sec:conclusion}

In this work we make the case that the design of permissioned blockchains should be revisited for datacenter-like environments and that latency should be considered as a first-class performance metric. With \emph{StreamChain}, we propose a streaming design and show that it achieves low latency while maintaining high throughput. We demonstrate that StreamChain could be already practical in real-world deployments with a supply chain management benchmark.

Our approach is complementary to ideas for further increasing throughput and, thanks to the elimination of batching overhead, novel research directions open up for hardware-accelerated consensus and cryptography operations in the context of permissioned ledgers. StreamChain is able to take advantage of low latency networking and modern datacenter hardware, including emerging platforms such as FPGAs, which, until now, have not been beneficial in such systems. 

\section*{Acknowledgments}
This project has received funding from the European Union’s Horizon 2020 research and innovation program under the Marie Skłodowska-Curie grant agreement No. 842956.

\balance

\bibliographystyle{abbrv}
\bibliography{full-streamchain}

\end{document}